# Photovoltaic Surfaces to Reverse Global Warming

Christiana Honsberg[1], Stuart G. Bowden[1], Ian R. Sellers[2], Richard R. King[1], Stephen M. Goodnick[1]

[1]Arizona State University, Tempe, AZ, 85281, USA
[2] Univeristy of Oklahoma, Norman, OK, 73019, USA

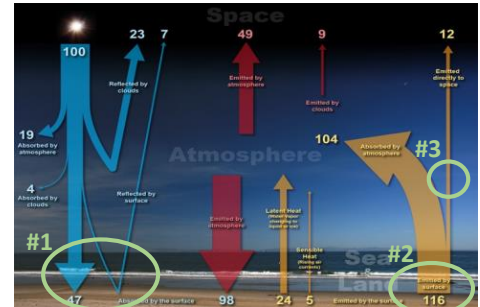

Figure 1: Earth's energy balance components

*ABSTRACT* — Climate changes and its many associated impacts are one of the most critical global challenges. Photovoltaics has been instrumental in mitigation of $CO_2$ through the generation of electricity. However, the goal of limiting global warming to 1.5 °C increasingly requires additional approaches. The paper presents how PV surfaces can be designed to reverse the Earth's radiative imbalance from increased greenhouse gasses that lead to higher global temperatures. The new PV surface generate electricity, reflect sub-band gap radiation, minimize their temperature, generate thermal radiation and emit additional IR through the atmospheric, with these processes totaling 650 W/m². This is realized by: (1) PV system efficiency at operating temperature > 20% and sub-band gap reflection of 150 W/m² for a total of 350 W/m²; (2) Thermally emitted radiation (radiative cooling) of 150 W/m²; and (3) Active IR emission through an atmospheric window at 1.5 μm of 150 W/m². With such PV surfaces, we show that 10 TW of installed PV can reverse global warming. Using PV to balance global temperatures introduces additional considerations for PV, focusing on high efficiency, particularly high efficiency at operating temperatures, radiative cooling, and new processes for 1.5 μm emission. We find that depending on their design, PV panels can increase or decrease global temperatures.

I. CORRECTING THE EARTH'S RADIATIVE BALANCE

The Earth's temperature is a result of the balance between incoming and outgoing radiation, shown in Figure 1. While the atmosphere and cloud cover have a large effect, the processes at the Earth's surface are also important. The Earth's surface absorbs about half (45 – 52% depending on data sources) of the solar radiation incident on it and 7% (4% –10%) is reflected and passes through the atmosphere. For these calculations, we need to differentiate between the ground surface of the Earth and the Earth's atmosphere, and clouds account reflect 23% of incident radiation into space.

In addition to absorbing and reflecting sunlight, the Earth's surface also emits radiation based on its temperature. Most of this emitted radiation is re-absorbed by the atmosphere (89% in Figure 1) and the remaining 11% is emitted through the atmospheric window into space. The atmosphere re-emits the absorbed radiation into space (47%) and towards the Earth's surface. Atmospheric conduction, convection, and latent heat also affect the Earth's energy budget, but PV surfaces are unlikely to have a significant effect on these processes and we do not further consider them. Overall, the processes into and out of the atmosphere equal, as shown by the numbers at the top of Figure 1.

Increasing greenhouse gasses in the atmosphere alters the radiation balance through multiple channels, including changing cloud cover/reflectivity and changing absorption of the atmosphere. An estimate of the imbalance in the incoming and outgoing radiation due to greenhouse gasses is 0.47 – 0.87 W/m² [ 1 ]. This imbalance in the Earth's radiative processes causes an increase in global temperatures.

The increase in the Earth's temperature can be mitigated or reversed by several approaches. PV reduces the generation of greenhouse gasses. However, it is increasingly difficult to reduce $CO_2$ sufficiently in a timeframe that limits global temperature rise to 1.5 °C, and so climate adaptation and climate engineering approaches are of increasing interest [ 2 ]. While these may become tools in managing the Earth's temperature, they presently suffer from issues ranging from economic or technical feasibility; reversibility and unintended consequences [3]; and scale necessary to achieve significant impact.

This paper demonstrates a new way to reverse the rising temperatures by designing photovoltaic surfaces to provide a global cooling effect by increasing the global radiative emission by 0.47 – 0.87 W/m² such that the Earth's temperatures are balanced even with additional greenhouse gasses and the Earth's temperature does not rise. The key requirements to control global temperatures from PV surfaces are labelled in Figure 1 and are: (1) Increasing the effective albedo of PV through generation of electricity and sub-band gap reflection; (2) Maximizing the thermal radiative emission of the PV modules; (3) Emitting 150 – 300 W/m² into space through an atmospheric window in near IR, targeting 1.5 μm. The following sections discuss these three components and the numbers on how to achieve them, showing that 10 TW of PV surfaces with effective emissions of 650 W/m² can balance the Earth's temperature.

**The abstract does not give details on the conventional PV model components; the detailed balance, optical, temperature, solar cell device and PV systems models have been used and described by the authors in other contexts.**

## II. COMPONENT 1: INCREASE EFFECTIVE ALBEDO OF PV

The first requirement for PV surfaces to provide net cooling is that they must have an effective albedo equal or greater than the surroundings and/ or the global average – i.e., it must not generate more heat than its surrounding surfaces. The Earth's albedo varies significantly between different types of surfaces and time of year, but a typical albedo of the land area relevant for PV is 30 – 40% (Figure 2a).

For a PV system, we define the effective albedo as:

$$\alpha_{eff} = \alpha(\lambda) + \eta_{OT} + \frac{Radiative\ recomb}{Incident\ radiation} \quad (1)$$

where $\alpha_{eff}$ is the effective albedo; $\alpha$ is the reflectivity of the photovoltaic surface (the albedo as typically defined); and $\eta_{OT}$ is the PV efficiency at operating temperatures (OT). The third term accounts for radiation emitted due to radiative recombination; for Si PV, we ignore this term since its maximum is 1 – 2%.

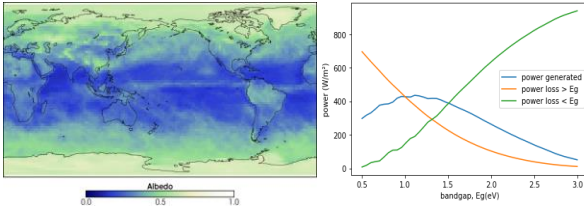

*Figure 2: (a) Earth's albedo. Calculation of albedo components for a solar cell as a function of band gap; power generation (blue) and sub-band gap reflection (orange).*

For an effective albedo of 30 – 40%, the power density from reflection and electrical generation is 300 – 400 W/m². Present PV systems have efficiencies of 15 – 20%, giving 150 – 200 W/m² of electrical power. Reflecting sub-band gap light increases the effective albedo, from a small value (~4% for glass PV surfaces) to a maximum of 170 W/m² for Si PV (Figure 2b). Sub-band gap reflection is incorporated by modifying the reflection properties of the front or adding reflectors at the rear [4].

Using 150 W/m² for the PV system power and 127 W/m² for sub-band gap reflection (75% reflection), existing PV systems have an effective albedo of 27.8%. Increasing PV system efficiency and the fraction of reflected sub-band gap light to 20% and 80%, gives an effective albedo of 33.6%. Optimization of the solar cell and optics shows that a thin Si solar cell (50 μm) and optimized reflection allows power densities of 410 W/m² while maintaining the yearly energy output. Overall, the consideration of radiative balance for PV systems introduces an additional advantage for high efficiency PV systems, particularly high efficiency at OT, as well as reducing the operating temperature of modules.

## III. COMPONENT 2: RADIATIVE COOLING

A second requirement for the PV surface is that the thermal emission power density equals or exceed that from the Earth's surface. This emission is often referred to as radiative cooling and is becoming of increasing interest for solar cells, night power generation, and in buildings [5,6]. For solar cells, radiative cooling combines with sub-band gap reflection and includes low visible reflection and high sub-band gap reflection while maximizing thermal emissivity.

The atmospheric window for radiative cooling is shown in Figure 4. The atmospheric window for thermally emitted radiation is from 5 – 12 μm and aligns with the peak black body (BB) radiation at 300K (orange line). The net 300K BB radiation emitted through the window is shown in Figure 3a (orange) and the integrated power density as a function of temperature is shown in Figure 3b. Figure 3b shows that increasing the temperature over the expected temperature range of a PV module changes the emitted radiation, with the increasing radiation at higher temperatures tempered by moving the BB radiation out of the ideal atmospheric window. The calculations are for an emissivity of 1, and we show (details in final paper) that etendue considerations minimize the impact of surface texturing on increasing thermal emission for optimally oriented surfaces.

Using the model for thermal emission, we calculate the power emitted from a PV system in Phoenix including the effect of module operating temperatures. We show that the PV radiative cooling can exceed that of local surroundings, exceeding the minimum required vale of 150 W/m². These results are corroborated by regional weather/climate modelling [7], which shows that higher daytime temperatures are offset by increased night radiation, leading to a net cooling effect.

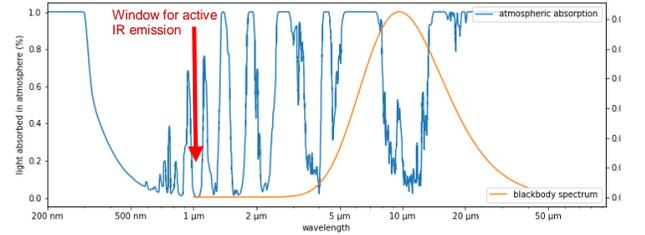

*Figure 4: Absorptance of the atmospheric (1 is completely absorbing) with a BB spectrum at 300K shown for reference. The window for active IR emission is shown by a red arrow.*

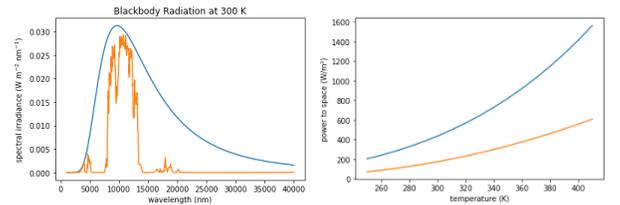

*Figure 3: (a) Emitted BB radiation through window (orange) and BB radiation. (b) Integrated power density from BB radiation (blue) as a function of T and emitted through atmospheric window (orange).*

## IV. COMPONENT 3: ACTIVE IR EMISSION

The final component for balancing the Earth's temperature is active IR emission through atmospheric windows in the near IR, with an optimal window at 1.5 microns or 0.87 eV (Figure 4, red arrow). The IR emission has properties like an LED rather than a BB, enabling the use of narrow atmospheric windows in the near IR. The first part of this section covers the required power density to balance global temperatures; and the second briefly overviews approaches to achieving active IR emission.

The power density needed from the PV panels depends on the required global average radiative emission density to balance the Earth's temperature (0.47 - 0.87 W/m²), the surface area of the Earth, and the surface area of PV. We use an intermediate value of $\Phi = 0.6$ W/m² to balance global temperatures. The equation is:

$$H_{Active\ IR}\left(\frac{W}{m^2}\right) = \phi_{global\ Avg}\left(\frac{W}{m^2}\right) \times \frac{Earth\ area}{PV\ pannel\ area} \quad (2)$$

The Earth's surface area is $5.1 \times 10^8$ km². Estimates of amount of PV needed for high renewable futures vary over nearly two orders of magnitude [8] up to 100 TW [9]; We use 10 TW [10] and $1.1 \times 10^6$ km², giving 278 W/m² needed from active IR emission. This value depends on the other processes. Increasing solar cell efficiency, sub-band gap emission, and radiative cooling while lowering PV's operating temperatures can reduce the power density of active IR emission needed, although not in a one-to-one fashion.

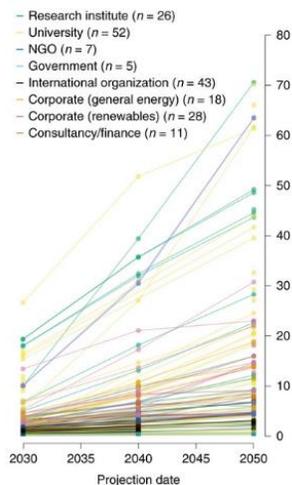

*Figure 5: Summary of non-IPCC solar energy scenarios [8].*

One approach to generating active IR emission is the use PV to power IR LEDs. The large power density, efficiency and availability of LEDs with appropriate emission wavelengths make this attractive. However, the total required electricity is large ($2.7 \times 10^8$ TWh) or 12 times the expected electricity use in 2035, and equal to the expected electricity use in 2050. The use of IR LEDs during curtailment of the PV system is a more realistic short-term approach; instead of generating heat during curtailment or operation away from maximum power point, the electricity powers IR emitters that reduce global radiation imbalances. Scenario calculations based on combining geographical regions into blocks with similar curtailment show that 13% of equivalent $CO_2$ can be removed using 10 TW of installed PV capacity which is optimized for components 1 and 2 as described in Sections II and III. The PV uses otherwise wasted electricity and removes an equivalent of 7 ppm of $CO_2$ from the atmosphere, or 19 GT of $CO_2$.

An alternate approach is to utilize the otherwise unused solar spectrum and convert it to near IR emission. Low energy (sub-band gap) light in the solar spectrum can be efficiently reflected into space since atmospheric scattering is minimized at these wavelengths. Hence, the goal is to utilize energy above the band gap more efficiently, shown in orange in Figure 2b, with nearly 400 W/m² available. There is a continuum of approaches ranging from technology-based (e.g, tandems) to utilizing new physical processes. The final paper shows using detailed balance models how advanced concepts such as hot carriers are important in realizing active IR emission and balancing the Earth's temperature.

## V. CONCLUSION

Photovoltaic surfaces can be designed to reverse global temperature rise. The requirements for the PV surfaces are to radiate through the atmospheric window and generate electricity with a total value of $> 650$ W/m². This can be achieved by: (1) increasing PV efficiency at operating temperatures (200 W/m²) and increasing sub-band gap reflection (150 W/m²); (2) increasing thermal IR emission to close to its maximum value (150 W/m²); and (3) and adding active IR emission in near IR (1.5 μm) of 150W/m².